\documentclass[a4paper, amsfonts, amssymb, amsmath, reprint, showkeys, nofootinbib, twoside]{revtex4-1}
\usepackage[english]{babel}
\usepackage[utf8]{inputenc}
\usepackage[colorinlistoftodos, color=green!40, prependcaption]{todonotes}
\usepackage[pdftex, pdftitle={Article}, pdfauthor={Author}]{hyperref} 
\begin{document}
\title{Resonance for life: Metabolism and Social Interactions in Bacterial Communities}

\author{Eleonora Alfinito$^{a}$, Matteo Beccaria$^{a,b,c}$}
    \email[Correspondence email address: ]{eleonora.alfinito@unisalento.it}
    \affiliation{${}^a$Department of Mathematics and Physics ‘Ennio De Giorgi’, University of Salento, I-73100 Lecce, Italy}
    \affiliation{${}^b$National Institute for Nuclear Physics (INFN) Sezione di Lecce, Via Arnesano, I-73100 Lecce, Italy}
    \affiliation{${}^c$National Biodiversity Future Center, Palermo 90133, Italy}

\date{\today} 

\begin{abstract}
The social organization of microorganisms has long been a fascinating and challenging subject in both biology and sociology. In these organisms, the role of the individual is far less dominant than that of the community, which functions as a superorganism. The coordination is achieved through a communication mechanism known as quorum sensing. When the community is healthy, quorum sensing enables it to regulate the development of potentially harmful individuals. This study employs an agent-based quorum sensing model to explore the relationship between metabolic functions and social behavior. It also examines how a polyculture responds to variations in the metabolic characteristics of its components. Finally, we identify a particularly stable condition in which the components cooperate to maximize the overall health of the colony. We refer to this state as  {\it  resonance for life}

\end{abstract}


\maketitle
\section{Introduction}  
The social behavior of microbes presents many fascinating aspects and has become the focus of increasingly in-depth studies. It is no coincidence that the term sociomicrobiology was coined in 2005 by Parsek and Greenberg [1]  as the “investigation of any group-behaviors of microbes”.
Social behavior implies an individual and a society as distinct entities. However, this is the first concept to be questioned in this field: how do we define an individual if it lacks self-awareness? And what constitutes conscious behavior at this level?
Cooperation, mutualism, cheating are behaviors we typically associate with evolved organisms from an anthropocentric perspective. We often perceive these beings as entities capable of making individual choices about their destiny. However, a more objective study – particularly of bacteria – suggests that the colony, rather than the individual, is the key functional. The colony acts as a kind of "super-individual", sustaining its components, producing defenses against enemies, mitigating defects, and acquiring resources. All this occurs without conscious awareness by instead through genetic programming and communication mechanisms that regulate gene expression across vast distances within the colony. As the theory of complex networks suggests, these mechanisms must be hierarchical [2,3] to ensure an effective response to external threats, preserving the colony’s deeper structural integrity.
Essentially, the communication mechanism does not induce cooperation in a social sense but rather facilitates structural and functional aggregation. In bacterial studies, this process is known as quorum sensing (QS), as it was first identified as the mechanism by which bacteria detect their population density – likely distinguishing between siblings and non-siblings [4-6]. Based on this count, the system can grow, produce public goods, protect itself against enemies or adverse environmental conditions [7-11] and, as various studies suggest [12-16], regulate behavioral anomalies.
In a study conducted a few years ago, Bruger and collaborators [12-14] analyzed the behavior of different mutants of {\it V. harveyi}, a bacterium well known for its bioluminescence (most famously observed in milky seas phenomenon [17] ). Due to its widespread presence, it serves as a model organism for bacteria studies. In particular, the authors of [12] engineered mutants with defects in the QS circuit. They examined the natural wild-type (WT) strain (BB120) alongside two mutants: one lacking the {\it luxR} gene which encodes the master regulator LuxR ({\it $\Delta$luxR}), and another deprived of the {\it luxO} and {\it luxU} genes. The first mutant, which reproduces poorly and does not produce public goods (PG), is classified as a defector. The second mutant, however, is termed an unconditional cooperator (UC) because it prioritizes PG production over offspring generation. The WT strain is able to counteract the growth of both defectors and UCs, whereas UCs are easily outcompeted by defectors. The authors of [12] concluded that QS – fully functional only in the WT – is the key mechanism enabling it to regulate the spread of both mutant types.
In this paper, using a previously developed QS model, we draw inspiration from the studies conducted by [12] to explore a possible metabolic framework for social behaviors. Specifically, we propose a metabolism-based behavioral phase diagram, i.e. a continuous behavioral landscape within which to identify the most common social traits observed in bacteria. By analyzing this diagram alongside the public goods (PGs) produced by an ideal colony, we also investigate the competitive dynamics of various social types. Finally, we highlight how competition between organisms with vastly different social behaviors can lead to a condition of serendipity, where both contenders maximize their outcomes by utilizing each other's resources.
\section{Materials and Methods}
\subsection{Materials}
As mentioned in the Introduction, quorum sensing (QS) is a coordination mechanism among bacteria mediated by the production and detection of signaling molecules called autoinducers (AIs)[ 5]. A functional QS circuit equips each bacterium with genes encoding both AIs and their corresponding receptors. Once released into the environment, AIs can be detected by all bacteria capable of sensing them, potentially over any distance. However, coordination effects only become significant when the bacterial population reaches a critical size. To replicate this phenomenon, we introduced a long-range communication mechanism driven by the number of bacteria.
The model begins with an initial cluster of agents within a confined space, where positions are evenly distributed in a regular pattern (grid). Each agent can occupy only a single position, and each position can host only one agent.
Each agent is defined by a set of parameters that determine its phenotype and sensing charge, $Q$. Here, $Q$ represents the source of interactions between agents. While real bacteria possess a surface electric charge [18], $Q$ does not necessarily describe this property. Instead, in this context, it is more closely related to the agent's size. Thus, each agent should be considered an aggregate of bacteria. Initially, agents are randomly distributed in a two-dimensional space with $Q$ set to 1. They then spread or multiply based on specific metabolic characteristics.
The metabolic features that characterize a single agent includes the maximum lifespan, ${\tau}_{max}$, measured in iteration steps and set to 10; the minimum size required for reproduction, $Q_{min}$, set to 2; the assimilation rate, $\sigma$; and the productivity index,$\alpha$. Each agent has an internal clock (${\tau}$) that increments with each iteration in which reproduction does not occur $\tau \rightarrow$ $\tau$+1). When an agent reproduces, its aging resets $\tau \rightarrow$ 0). If $\tau$ reaches ${\tau}_{max}$, the agent disappears.
\subsection{Methods}
The analysis is conducted at a coarse-grained level using a system of agents that move freely on a square grid, with movement restricted to the eight nearest neighbors. In addition to moving, agents can replicate, with the duplicate occupying one of the adjacent positions.
The study begins with an initial cluster of agents randomly distributed on the grid. Depending on their specific metabolic characteristics, they can either reproduce until the space is completely filled or continue exploring without successfully reproducing. In the first scenario, colony growth halts when resources are depleted, while in the second, agents gradually die of old age. The outcome is determined by metabolic parameters, allowing for an exploration of parameter space to study different evolutionary trajectories. Specifically, we analyze growth performance as a function of two key parameters: assimilation rate and productivity index, while keeping other parameters constant.
The model described here applies to an ideal bacterial colony implementing quorum sensing (QS). It has previously been used to describe bioluminescence in {\it V. harveyi } and its dependence on three different types of autoinducers [19-21], but the results are broadly applicable to any form of bacterial growth.
The stochastic procedure follows these five steps:
\begin{enumerate}
\item Calculation of the energy E, and potential, V, of each agent. It is made according to the formula:
\begin{equation}
E_{a}=Q_{a} \sum_{b \ne a} \frac{Q_{b}}{D_{a,b}} = V_{a}Q_{a}
\end{equation}
where $D_{a,b}$   is the geometrical distance between the nodes ${a,b}$ .
\item Establishing connections between nodes: Each agent creates a link with other agents that have a lower potential. This step constructs the network.
\item Distribution of resources (sensing charges) among agents. This occurs with a higher probability the higher the total energy, the closer the energy of the agents involved and the lower the productivity index. Through the links, resources (sensing charges) are transferred and assimilated with the assigned assimilation rate $\sigma$ ($\sigma$ is in the range [0 – 1]). 
\item Reproduction/migration. Each agent with charge higher than $Q_{min}$ divides in equal parts (binary fission) and the daughter agent occupies the lowest potential position among the 8 nearest neighbors. Agents with charge lower than $Q_{min}$ moves to the lowest potential position among the 8 nearest neighbors. A larger than 1 value of the ${fitness}$  of the colony indicates that the initial nucleus of agents is grown [21].
\item Production of public goods. When a link is established, it is assigned an effective resistance. This resistance measures the difficulty of transferring information between agents and depends on factors such as distance and the propensity to produce public goods; it increases as this propensity decreases. Additionally, the resistance value decreases as the amount of charge distributed across the landscape and the productivity index increase.
An ideal pair of contacts is positioned at the ends of the grid, and a potential difference is applied across them. The resulting current serves as a measure of the network's connectivity and the amount of charge present in the landscape. We associate this current with the quantity of public goods that the colony is capable of producing [19].
The single resistance $r_{a,b}$ is given by the formula
\begin{equation}
r_{a,b}=\frac{D_{a,b}}{\alpha} \left [ r_{0} (1-h_{a,b})+ r_{1} h_{a,b} \right]
\end{equation}

where $\frac{r_{0}}{r_{1}}  =1000 $is an arbitrary input variable, and $h_{a,b}$ is a sigmoidal function of the sensing charge whose codomain is the range [0,1] [20].
\end{enumerate}
Simulations are performed on a 20x20 grid, and results are mediated over 60 realizations.
\section{Results}
\subsection{Metabolic origin of the social beaviors}
In the model, the assimilation rate ($\sigma$) and the productivity index ($\alpha$) compete in the distribution of charges to individual agents, influencing their development. By focusing on these two parameters, we created a growth phase diagram, which represents the probability of colony growth as a function of $\sigma$ and $\alpha$ . Moving from bottom to top or from right to left, the probability of reproduction decreases, eventually disappearing completely. In Figure 1, this behavior is depicted using a color scale from blue to red (low to high reproduction). The white region highlights the combinations of ($\sigma,\alpha$) that prevent colony development.
Within this diagram, we can identify various social categories, as suggested by studies like [10,12-14]. Dormants, for example, are agents that can only migrate. They move toward positions of minimum potential, where the likelihood of receiving charge is higher. This long lifespan state (determined by $\tau_{max}$) simulates the quiescent condition some microorganisms adopt when environmental conditions are too harsh.

\begin{figure}
    \centering
    \includegraphics[width=1.0\linewidth]{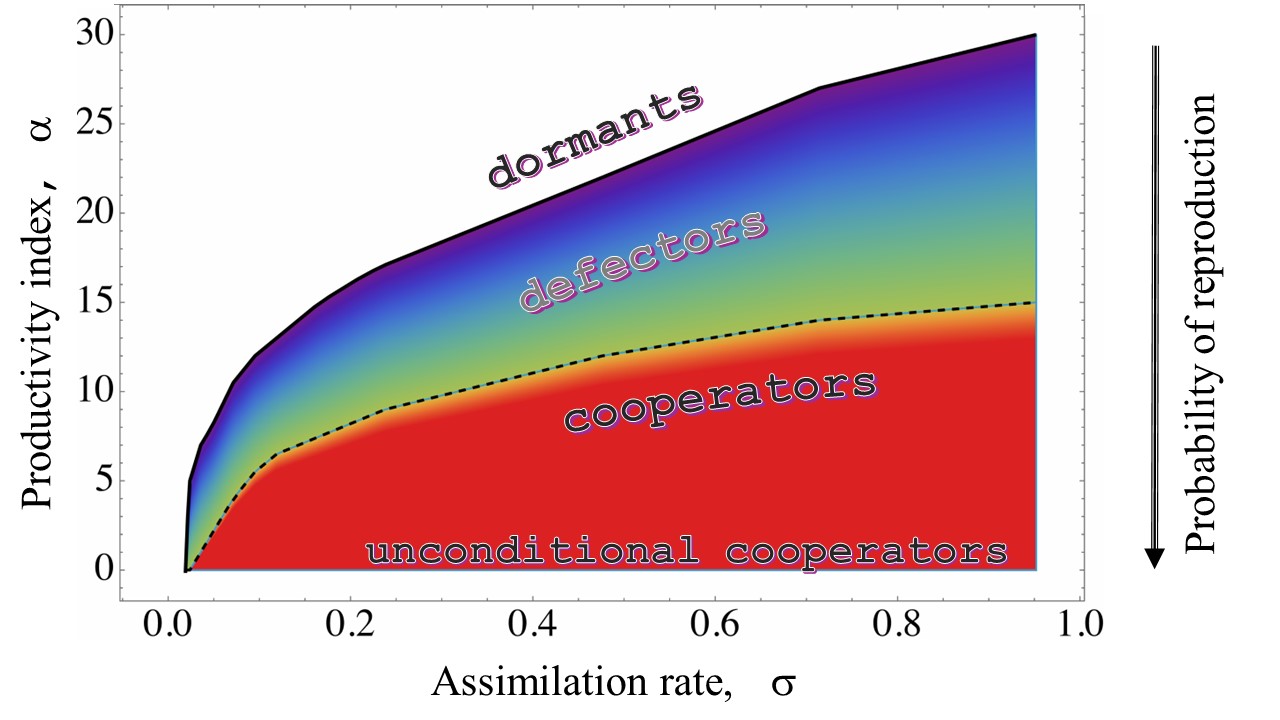}
    \caption{Reproduction phase diagram. Data reports the probability of reproduction as calculated by our simulations. The color scale goes from blue ( very low percentage of colonies able to reproduce ) to red (maximal probability of reproduction). White indicates that there is zero probability of reproduction. The dotted line highlights the conditions where the probability of reproduction is 97$\% $. Furthermore, the regions corresponding to the different social behaviors are indicated (dormants, defectors, cooperators, unconditional cooperators).}
    \label{fig:enter-label}
\end{figure} 
To better understand the behavior of the agents as the pair ($\sigma, \alpha$) varies, it is useful to examine the production of public goods (PG) by varying 
$\alpha$  or $\sigma$ individually. In Figure 2, we observe the system's response in terms of productivity, lifespan, and the percentage of ungrown colonies for varying $\sigma$ with a fixed productivity index ($\alpha$ = 10). It is important to note that varying $\sigma$ while keeping $\alpha$ constant is equivalent to drawing a horizontal line on the plot in Figure 1.
First, there is a minimum value of $\sigma$ (here,$\sigma$  = 9) below which colony development does not occur, and no PG production is achieved. As $\sigma$ increases, PG production reaches its maximum value, which is determined by the choice of $\alpha$. In other words, all initial configurations result in a fully developed and highly productive colony. However, the colony's survival time decreases as the metabolic rate increases, due to the rapid consumption of resources.
\begin{figure}
    \centering
    \includegraphics[width=1.0\linewidth]{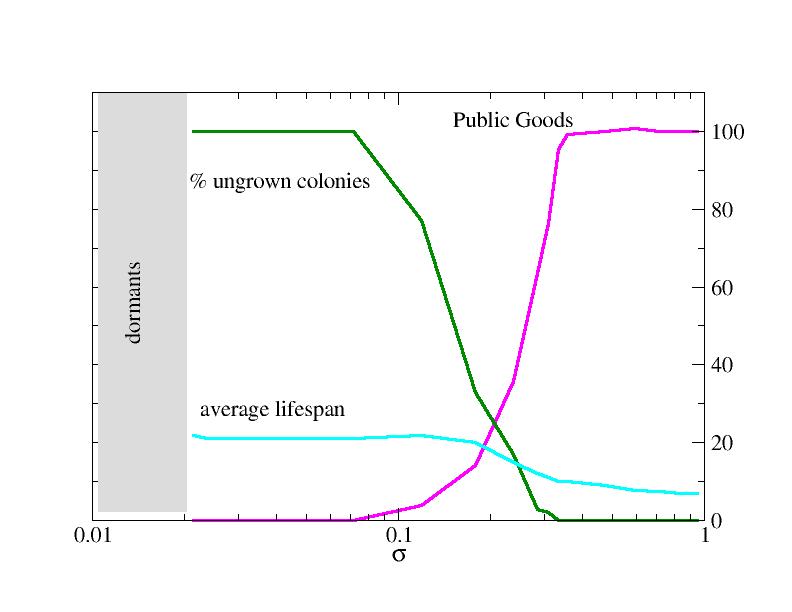}
    \caption{Colony development for assigned production index, $\alpha$ =10. The production of public goods, the percentage of ungrown colonies and the average lifespan are reported for different values of the assimilation rate, $\sigma$. }
    \label{fig:enter-label}
\end{figure}
In Figure 3, we show the effect of varying $\alpha$ on agents with a fixed $\sigma$ (represented by a vertical line in the phase diagram). In this case, the PG curve is bell-shaped, while both the average lifespan and the percentage of ungrown colonies maintain the saturation-like behavior observed in Figure 2. Specifically, as the demand for productivity increases, fewer colonies are able to develop, and their survival time increases because they consume fewer resources.
Figures 1-3 provide an interpretative framework for understanding social behaviors in metabolic terms. Specifically, we can consider the ideal operating condition (wild type) as the one in which both $\sigma$ and $\alpha$  are calibrated to maximize PG production and the ability to produce offspring. This condition corresponds to the central region of the PG curve in Figure 3, just to fix the ideas, we could identify this as the region between 60 $\%$  and 100$\% $ of PG production, generally identifying this social behavior as that of cooperators.
To the left of this region, we find colonies that develop rapidly, producing the maximum amount of PG required by the productivity index. These colonies can be identified as \textit{unconditional cooperators} (UC). To the right of this region, we find colonies that develop more slowly, eventually leading to colonies with a low probability of development. In each case, the PG yield is lower than the amount required by the productivity index. These colonies are identified as \textit{defectors} [12].
In this representation, entire regions of ($\alpha$, $\sigma$) combinations correspond to distinct social behaviors, each described with varying nuances. Inside these regions we can identify the specific features identified in [12]. For example, in [12], it is stated that $\Delta$\textit{luxR} reproduces at a rate of about 3$\%$ of the wild type (WT). Referring to Figure 3, this corresponds, approximately, to the pair ($\alpha$ = 20, $\sigma$ = 0.7), while the WT, represented by a single pair, would be ($\alpha$ = 7, $\sigma$ = 0.7).
\begin{figure}
    \centering
    \includegraphics[width=1\linewidth]{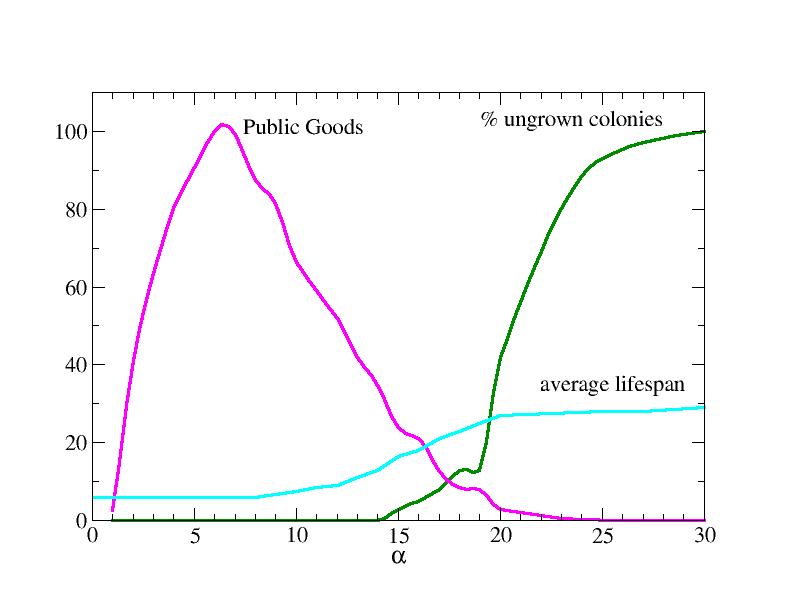}
    \caption{Colony development for assigned assimilation rate, $\sigma$=0.7. The production of public goods, the percentage of ungrown colonies and the average lifespan are reported for different values of the production index, $\alpha$}
    \label{fig:enter-label}
\end{figure}
In Figure 1, we have qualitatively indicated these regions, in other terms, the growth phase diagram can be also read as a behavioral ( social) phase diagram.
\subsubsection{3.2	QS mediates between different behaviors}
As highlighted in some papers [22-26], quorum sensing (QS) plays a crucial role in managing the well-being of a bacterial colony, particularly by mediating between different behaviors that could potentially harm it. In this model, we have interpreted this mediation in terms of the productivity index, $\alpha$. Specifically, when two different types of agents, specifically, with different $\alpha$ values, develop on the same landscape (grid), evolution occurs by preserving differences in assimilation rates while mediating the values of $\alpha$ . 
The ability to mediate between two different productivity demands creates a broad range of possible outcomes. To narrow this down, we will focus on two distinct cases: 1. agents with equal metabolism but different productivity, and 2. agents with equal productivity but different metabolism.
\begin{enumerate}
    \item In the case of two different types of agents with identical $\sigma$ values but different $\alpha$, common evolution leads to a single type of agent with production characteristics that are intermediate between the original types. This should represent a new, homogeneous colony capable of mimicking, at least qualitatively, the competition results observed in [12], where defectors and unconditional cooperators may invade the wild type (WT) but not overcome it, while defectors lead to the extinction of UCs. In our simulations, averaging over $\alpha$ shifts both UCs and defectors toward the central region when they compete with WT. Furthermore, since the defector region is much larger than that of UC, the resulting individual is closer to the defector, with a reduced ability to produce PG.
Smaller values of $\alpha$ produce very similar results, with the maximum PG production shifting to the smallest $\sigma$ values. This type of investigation concerns mutants of the same species (with identical assimilation rates).
As a final comment, this result is reminiscent of horizontal gene transfer, a phenomenon common in bacteria and of significant clinical interest, as it underpins the development of antibiotic resistance [27].
\item  A more general case study would involve competition between agents with all metabolic parameters differing. This creates a very broad range of possible outcomes, so we choose to focus on the case of agents differing in assimilation rate, selecting one of the competitors with the highest $\sigma$ value. Regarding the productivity index, because the model averages over different values, for simplicity, we analyze pairs of competitors with the same $\alpha$ value. For each value of $\alpha$, exploration follows a horizontal line in the phase diagram, covering multiple "social" regions. To clarify the discussion, we will refer to the competitor with the highest assimilation rate as the "host" and the other as the "intruder," even though both initially have the same concentrations.
As a general result, for each value of $\alpha$, the interaction between the two competitors exhibits different phases. 
\end{enumerate}
\begin{itemize}
    \item 
\textbf{Phase 1} The intruder has a very low assimilation rate and cannot spread, while the host quickly fills all the empty positions. The average lifespan of the colony is close to that of the host and much shorter than that of the intruder.
\item
\textbf{Phase 2}: The intruder has an intermediate assimilation rate and begins to spread (fitness between 1-2). Note that an intruder with these metabolic parameters, when grown in monoculture, is in a dormant state. However, in polyculture, due to the high number of charges produced by the host, the intruder can reproduce. This moderate competition allows the host to slow down its growth, resulting in a colony with a relatively high lifespan, longer than that of either the host or the intruder in monoculture: we call this condition \textit{resonance for life}, because it represents a condition of perfect coexistence of the two species.
\item\textbf{Phase 3}:As the assimilation rate of the intruder increases, it becomes capable of competing on equal terms with the host, increasing its fitness. This, in turn, reduces the colony's lifespan due to faster resource consumption.
\item\textbf{Phase 4}:Both the host and the intruders develop in the landscape with equal capacity, and the lifespan of both competitors tends to converge to the same value.
\end{itemize}
In Figure 4, we present this information in terms of normalized mean lifespans of the colonies. Normalization is performed by comparing the lifespan of the mixed colony to that of the intruder (which has a longer lifespan than the host). This allows us to assess the gain in lifespan for the mixed colony. The data were calculated for three different values of the productivity index: $\alpha = 10^4$, where the host is in the UC state; $\alpha$= 10, where it is in the cooperative (WT) state; and $\alpha$ = 20, where it is in the defector state. The behaviors are very similar, and to make the figure more readable, only the initial part of Phase 4 is reported (where the curve tends to 1).
\begin{figure}
    \centering
    \includegraphics[width=1\linewidth]{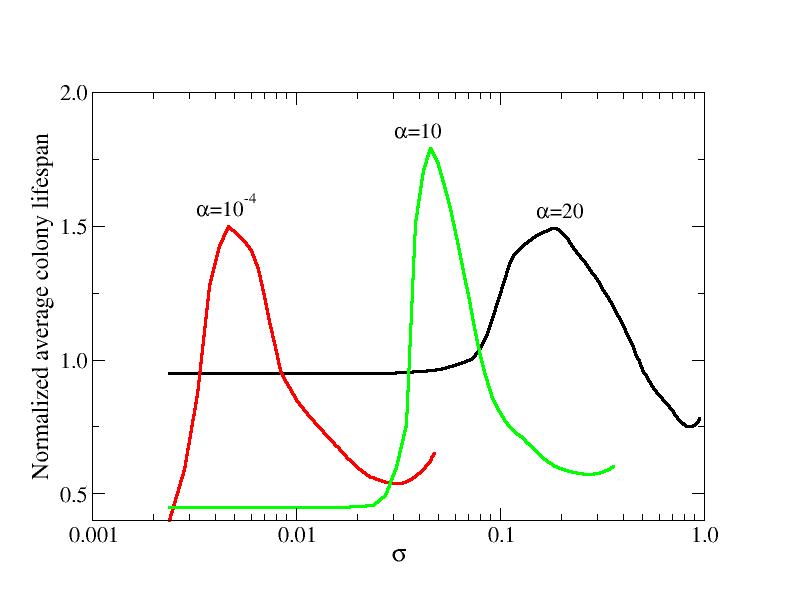}
    \caption{Resonance for life. The gain in lifespan obtained comparing the lifespan of the polyculture with that of the intruder monoculture. Phases 1,2,3 are reported (see text). }
    \label{fig:enter-label}
\end{figure}
\section{Discussion and Conclusions}
Quorum sensing (QS), a phenomenon discovered about 50 years ago [4-6], seems to replace the concept of self-awareness in bacteria with a collective, rather than individual, awareness. It develops alongside the colony and is an expression of it, as it manifests with the increase in colony size. At the same time, it serves as the colony's generator, overseeing cell growth and the production of public goods.
Therefore, having a theoretical model of QS can be a valuable tool for predicting the behavior of bacterial colonies in all aspects of their evolution. As highlighted in [28], a general theoretical model may be imprecise or overly broad unless its parameters are finely tuned to align with the specific biological model under study. In the present research, social behaviors and metabolic characteristics of a generic bacterial system are correlated, yielding a continuous spectrum of correspondences. Within this spectrum, we can also identify several social behaviors reported in the literature.
Finally, this model allows for the exploration of polycultures composed of two different types of bacteria. The results show, as a particular case, the formation of a monoculture with intermediate characteristics between the two initial types, similar to what occurs in horizontal gene transfer. Another significant finding is the emergence of a condition where both competitors derive an advantage (in terms of fitness or lifespan) from the presence of the other. This condition arises only when specific combinations of metabolic parameters are met.
As a final comment, we emphasize the extraordinary ability of microorganisms to organize themselves into highly efficient structures, capable of reacting to changing environmental conditions and seizing opportunities to exploit or contain elements with different metabolic characteristics. Any model or prediction that can be developed is ultimately confirmed by what natural evolution has already explored and refined. This, too, is not surprising; it is, in fact, one of the fundamental principles of Darwin's theory of \textit{the struggle for life}.

\section*{References}
\begin{enumerate}
    \item 
Parsek, M. R.; Greenberg, E. P. Sociomicrobiology: the connections between quorum sensing and biofilms. Trends Microbiol 2005,13(1), 27-33.
\item 
Ravasz, E.; Barabási, A. L. Hierarchical organization in complex networks. Phys Rev E 2003, 67(2), 026112.
\item Redhead, D.; Power, E. A. Social hierarchies and social networks in humans. Philos Trans R S B 2022, 377(1845), 20200440.
\item Nealson K.H.; Hastings J.W. Bacterial bioluminescence: its control and ecological significance. Microbiol Rev 1979, 43 , 496–518.
\item Miller, M. B.; Bassler, B. L. Quorum sensing in bacteria. Annu Rev Microbiol 2001 , 55(1), 165-199.
\item Bassler, B. L.; Losick, R. Bacterially speaking. Cell 2006 ,125(2), 237-246.
\item Abisado, R. G.; Benomar, S.; Klaus, J. R.; Dandekar, A. A.;  Chandler, J. R. Bacterial quorum sensing and microbial community interactions. MBio 2008, 9(3), 10-1128.
\item Ayrapetyan, M.; Williams, T. C.; Oliver, J. D. Interspecific quorum sensing mediates the resuscitation of viable but noncul-turable vibrios. Appl Environ Microbiol 2014, 80(8), 2478-2483.
\item Bari, S. N.; Roky, M. K.; Mohiuddin, M.; Kamruzzaman, M.; Mekalanos, J. J.; Faruque, S. M. Quorum-sensing autoinducers resuscitate dormant Vibrio cholerae in environmental water samples. PNAS 2013, 110(24), 9926-9931.
\item Personnic, N.; Striednig, B.; Hilbi, H. (2021). Quorum sensing controls persistence, resuscitation, and virulence of Legionella subpopulations in biofilms. ISME J 2021, 15(1), 196-210.
\item Dandekar, A. A.; Chugani, S.; Greenberg, E. P. Bacterial quorum sensing and metabolic incentives to cooperate. Science 2012, 338(6104), 264-266.
\item Bruger, E. L.; Waters, C. M. (2016). Bacterial quorum sensing stabilizes cooperation by optimizing growth strategies. Appl En-viron Microb 2016,  82(22), 6498-6506.
\item Bruger, E. L.; Waters, C. M. (2018). Maximizing growth yield and dispersal via quorum sensing promotes cooperation in Vib-rio bacteria. Appl Environ Microb 2018, 84(14), e00402-18.
\item Bruger, E. L.; Snyder, D. J.; Cooper, V. S.; Waters, C. M. Quorum sensing provides a molecular mechanism for evolution to tune and maintain investment in cooperation. ISME J 2021, 15(4), 1236-1247.
\item Smalley, N. E.; An, D.; Parsek, M. R.; Chandler, J. R.; Dandekar, A. A. (2015). Quorum sensing protects Pseudomonas aeru-ginosa against cheating by other species in a laboratory coculture model. J Bacteriol  2015, 197(19), 3154-3159.
\item Zhao, K.; Liu, L.; Chen, X.; Huang, T.; Du, L.; Lin, J.; ...  Chu, Y. Behavioral heterogeneity in quorum sensing can stabilize social cooperation in microbial populations. BMC biology, 2019, 17, 1-15.
\item
Miller, S. D.; Haddock, S. H.; Straka III, W. C.; Seaman, C. J.; Combs, C. L.: Wang, M.; ...  Nam, S. Honing in on biolumi-nescent milky seas from space. Sci Rep 2021, 11(1), 15443.
\item Wilhelm, M. J.; Gh, M. S.; Wu, T.; Li, Y.; Chang, C. M.; Ma, J.; and Dai, H. L. . Determination of bacterial surface charge den-sity via saturation of adsorbed ions. Biophys J 2021, 120(12), 2461-2470.
\item	Alfinito, E.; Cesaria, M.; Beccaria, M. Did Maxwell dream of electrical bacteria? Biophysica 2022, 2(3), 281-291.
\item Alfinito, E.; Beccaria, M.; Cesaria, M. Cooperation in bioluminescence: understanding the role of autoinducers by a stochastic random resistor model. EPJ E 2023,46(10), 94.
\item Alfinito, E.; Beccaria, M. (2024). Competitive Distribution of Public Goods: The Role of Quorum Sensing in the Development of Bacteria Colonies. Biophysica 2024, 4(3), 327-339.
\item Dudley, S. A. Discovering cooperative traits in crop plants. PLoS Biology 2022, 20(11), e3001892.
\item Henke, J. M.; Bassler, B. L. Bacterial social engagements. Trends Cell Biol 2004, 14(11), 648-656.
\item Bruger, E.; Waters, C. Sharing the sandbox: evolutionary mechanisms that maintain bacterial cooperation. F1000Research 2015, 4. F1000-Faculty.
\item Ben-Jacob, E. (2008). Social behavior of bacteria: from physics to complex organization. EPJ B 2008, 65, 315-322.
\item Coyte, Katharine Z.; Rakoff-Nahoum, S. Understanding competition and cooperation within the mammalian gut microbi-ome. Curr Biol 2019 29(11), R538-R544.
\item Michaelis, C.,  Grohmann, E. Horizontal gene transfer of antibiotic resistance genes in biofilms. Antibiotics 2023, 12(2), 328.
\item	Fourcade, Y. Fine-tuning niche models matters in invasion ecology. A lesson from the land planarian Obama nungara. Ecol Model 2021, 457, 109686.
\end{enumerate}
\end{document}